\documentclass[11pt,twoside]{article} 
\usepackage{asp2004}
\usepackage{epsf}
\usepackage{psfig}
\usepackage{lscape} 

\begin{document} 
\title{Discovery of magnetic fields in hot subdwarfs}
\author{S.~J. O'Toole$^1$, S. Jordan$^2$, S. Friedrich$^3$, and
U. Heber$^1$}
\affil{$^1$Dr Remeis-Sternwarte, Astronomisches Institut der
  Universit\"at Erlangen-N\"urnberg, Sternwartstr.\ 7, Bamberg D-96049,
  Germany \\
 $^2$Astronomisches Rechen-Institut, M\"onchhofstr.\ 12-14, 69120
  Heidelberg, Germany \\
$^3$Max-Planck-Institut f\"ur extraterrestrische Physik, Giessenbachstr.,
  85748 Garching, Germany}
\begin{abstract} 
We present initial results of a project to measure mean longitudinal magnetic
fields in a group of sdB/OB/O stars. The project was inspired by the discovery
of three super-metal-rich sdOB stars, each having metals (e.g. Ti, V) enhanced
by factors of 10$^3$ to 10$^5$. Similar behaviour is observed in chemically
peculiar A stars, where strong magnetic fields are responsible for the
enrichment. With this in mind, we obtained circularly polarised spectra of two
of the super-metal-rich sdOBs, two "normal" sdBs and two sdOs using FORS1 on
the ESO/VLT. By examining circular polarisation in the hydrogen Balmer lines
and in helium lines, we have detected magnetic fields with strengths of 1-2 kG
in most of our targets. This suggests that such fields are relatively common in
hot subdwarfs. 
\end{abstract}

\section{Motivation}
\par Recently, we have found several sdBs that show a very large enrichment of
iron group elements (Edelmann et al. 2001). Three out of thirteen targets
showed abundances of these elements of up to 10$^5$ times the solar
value. Around 23\% of the lines in the peculiar sdBs remain
unidentified. These metal-enriched sdBs are reminiscent of the chemically
peculiar A stars, where strong magnetic fields play a major role in the
surface abundance patterns. Although no direct evidence for magnetic fields
was found, we felt a proper examination of these peculiar objects for such a
field was clearly required.

\par Strittmatter \& Norris (1971) studied the role of magnetic fields in
chemically peculiar A stars, and several of their conclusions may be relevant
to sdBs. The abundance pecularities in Ap stars are, in general,
confined to the stellar surface. If the magnetic field is strong enough,
Strittmatter \& Norris suggest that magnetic braking will occur, causing
angular momentum loss and resulting in a slowly rotating star. The slowing of
the star's rotation can lead to peculiar surface abundances, however there is
no well established or detailed theory. In the case of sdB stars, the vast
majority are slowly rotating, with observational limits of $v\sin
i<5$\,km\,s$^{-1}$. The only sdB star for which rotation has been detected is
PG1605+072 (Heber et al. 1999). The metal lines of all the stars studied by
Edelmann et al. (2001) are very sharp, consistent with
projected rotational velocities of $v\sin i<5$\,km\,s$^{-1}$. Could
the slow rotation of sdB stars, coupled with a magnetic field of 1-2
kG, be the cause of the peculiar abundance patterns of some sdB and
sdOB stars? To test this idea, we used the ESO/VLT equipped with FORS1
to investigate circular polarisation in hot subdwarfs with peculiar
surface abundances. We have also observed apparently "normal" sdB and
sdO stars for comparison.

\par Elkin (1996) conducted a search for magnetic fields in sdB and sdO stars
using the 6m Special Astrophysical Observatory, however, his study used several
different measurement techniques and was therefore quite inhomogeneous. Our
study is the first homogeneous search for magnetic fields
in sdB stars. Elkin (1996) reported the detection of fields in two targets,
the sdO star BD\,$+75^\circ 325$ and the sdOB star Feige\,66. The other results
were presented in Elkin (1998) and were largely inconclusive, with errors for
some targets approaching $\pm$1 kG. We do not believe this is sufficient for
sdB stars, since, based on Elkin's detection of a variable field in the sdOB
star Feige 66 with extrema of $-$1.3 kG and +1.75 kG, we expect field strengths
to be no larger than $\sim$1-2 kG. By measuring the polarisation of the Balmer
lines using the method of Bagnulo et al. (2002), we expected our typical error
for each target would be 200-300 G.

\section{Observations}

\par Our observations were made with FORS1 at the VLT using the
spectropolarimetry setup discussed by Bagnulo et al. (2002). Briefly, a
Wollaston prism and a quarter-wave retarder plate were placed in the beam. We
used the 600B grism which allowed spectropolarimetry of the Balmer sequence
from H$\beta$ to the Balmer discontinuity, as well as several strong helium
lines. Exposures were taken using two orientations of the retarder plate
separated by 90$^\circ$ to cancel out instrumental polarisation.

\par Both of our
super metal-rich stars were observed, along with a "normal" sdB, an evolved,
post-EHB sdB, and two helium-rich sdOs. Feige 66, the sdOB star with a
previously measured magnetic field, was not observable due to weather
constraints.

\section{Results}

Previous studies of circularly polarised stellar spectra have focussed solely
on the Balmer lines (e.g. Bagnulo et al. 2002; Aznar-Cuadrado et al. 2004),
because they  are strong in Ap stars and DA white dwarfs. This is true of many
hot subdwarfs, but often helium lines are also present, and in some cases
hydrogen is extremely deficient. In order to realise the full potential of our
spectra, we calculated Lande $g$-factors for the strongest
He\,\textsc{i} and He\,\textsc{ii} lines.

\begin{figure}
\plotone{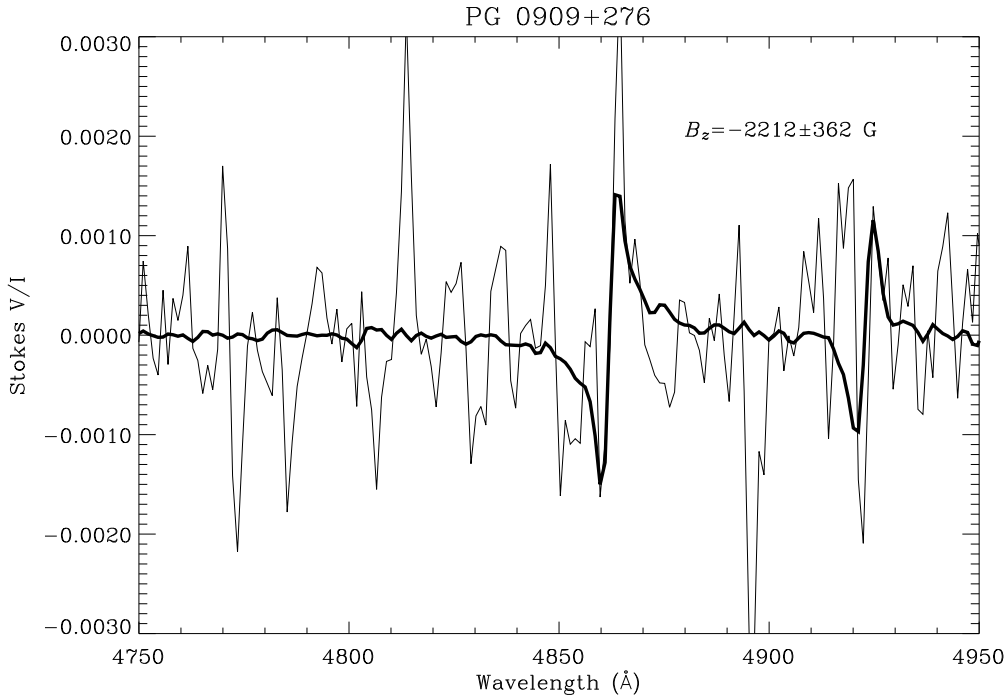}
\caption{Stokes \emph{V/I} spectrum of PG\,0909+276, centered on H$\beta$. The
  He\,\textsc{i} 4921\,\AA\ line is also shown.}
\label{fig:pg0909}
\end{figure}

\begin{figure}
\plottwo{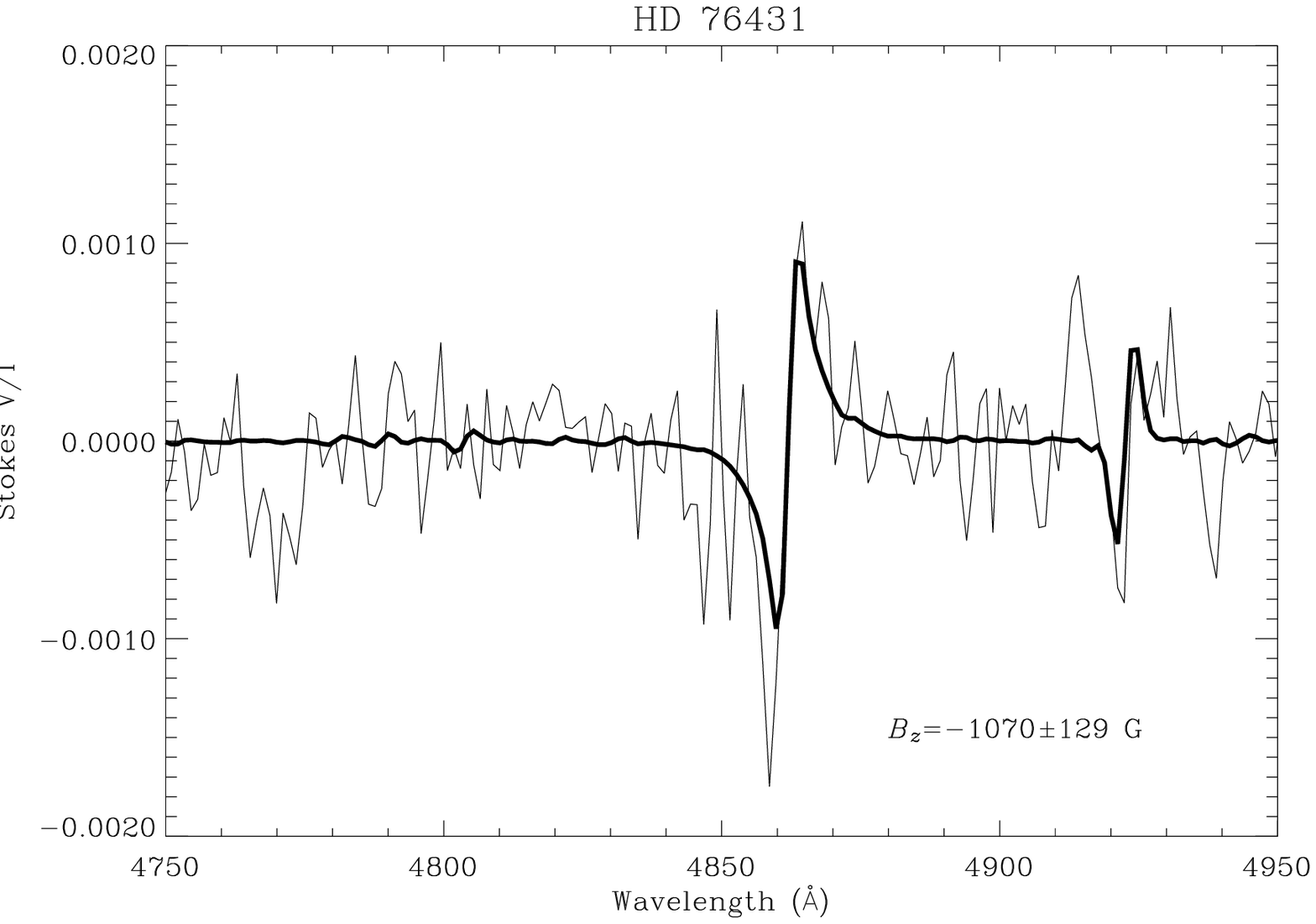}{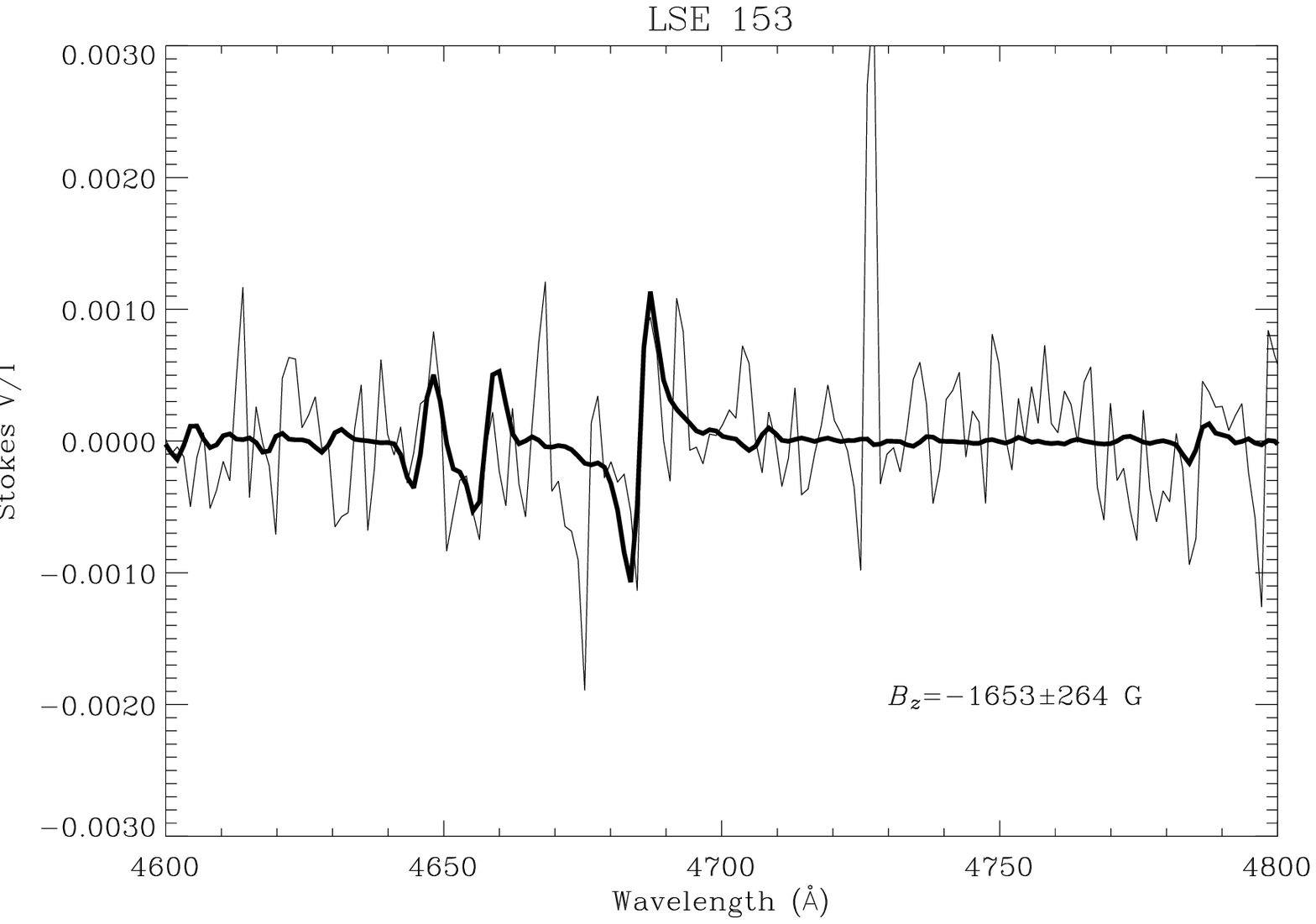}
\caption{Stokes \emph{V/I} spectra for HD\,76431 (\emph{left}) and LSE\,153
  (\emph{right}). As for PG\,0909+276, H$\beta$ and the He\,\textsc{i}
  4921\,\AA\ are shown. In the case of LSE\,153, the line displayed is
  He\,\textsc{ii} 4686\,\AA.}
\label{fig:lse153}
\end{figure}

The values for the two strongest lines are He\,\textsc{i} 4471\,\AA: 1.17; and
He\,\textsc{ii} 4686\,\AA: 1.07. Note that for all Balmer lines the Lande
g-factor is approximately unity. Comparing our measured Stokes
\emph{V/I}  profiles with those expected based on the weak-field
approximation, we were able to measure the magnetic field of each of
our targets. As in Aznar Cuadrado et al.\ (2004) we used $\chi^2$ fits
of the polarisation pattern in the vicinity of strong lines in order
to derive mean longitudinal magnetic fields and their respective error
margins. Observed profiles of three of our targets are shown in
Fig. \ref{fig:pg0909} and \ref{fig:lse153}, along with our best fit.

Using this method, we have made clear detections in five of six targets:
UVO\,0512$-$08, PG\,0909+276, HD\,76431, LSE\,153 and CPD\,$-64^\circ 481$,
with measured field strengths of $-1373\pm328$\,G, $-2212\pm362$\,G,
$-1070\pm129$\,G, $-1675\pm264$\,G and $-1208\pm266$\,G, respectively. These
values are weighted averages of several lines, the number of strong, usable
lines depending
on the helium abundance of the star. The sixth star in our sample,
CD\,$-31^\circ 4800$, shows a possibly significant field strength, however,
most lines in its spectrum are blended.

Since there is no clear difference between the field strengths of the "normal"
and super metal-rich stars in our sample, it is unlikely that magnetic fields
are directly connected to large abundance peculiarities in hot subdwarfs. The
origin of the fields detected here are unknown, and the fields are found in
objects from various evolutionary channels. While the two sdOBs and two sdBs
have probably followed similar evolutionary channels, the same cannot be said
for at least one of the sdO stars. LSE\,153 can be associated with the
"born-again" post-AGB scenario, while the helium-rich sdO, CD\,$-31^\circ
4800$, may not be related to the hydrogen rich sdB and sdOB stars. If we
assume that all of these objects will evolve into white dwarfs, then we can
estimate what the field strength will be if the magnetic flux is completely
conserved. The radii of the sdB/OB stars are 0.15-0.25\,$R_\odot$ (taking the
canonical mass of an EHB star to be 0.48\,$M_\odot$). Since they will evolve
directly into white dwarfs, their radii will shrink by
a factor of $\sim$20, leading to field strengths of up to 500kG, values which
are apparently rarely seen. The search for
rotation in white dwarfs (Heber et al.\ 1997; Koester et al.\ 1998; Karl et
al.\ 2004) from high resolution H$\alpha$ spectroscopy also resulted in
constraints on field strengths. Amongst the $\sim$50 DA white dwarfs studied,
only four turned out to be magnetic with $B$ up to 180\,kG, while upper limits
of 10-20\,kG could be derived for the rest. 

\section{Outlook}

The detection of significant polarisation in our sample of hot subdwarfs
suggests that magnetic field strengths of $\sim$1-2 kG are actually quite
common among these stars. It remains to be seen whether or not hot subdwarfs
all have such fields, or if for some reason our study simply "got lucky". With
this in mind, we plan to survey magnetic fields in
a larger number of targets.

\acknowledgements{SJOT is supported by the Deutsches Zentrum f\"ur Luft- und
Raumfahrt (DLR) through grant no.\ 50-OR-0202.}

\end{document}